\documentclass[authoryear]{elsarticle}
\usepackage{ifpdf}
\usepackage{graphicx,amssymb,amsmath,lineno}
\usepackage{enumitem}
\usepackage{bm}
\usepackage[percent]{overpic}
\usepackage{setspace}
\usepackage{scalerel}
\usepackage{url}
\usepackage{todo}
\usepackage{gensymb}

\ifpdf
\usepackage[%
  pdftitle={Instructions for use of the document class
    elsart},%
    pdfstartview=FitH,%
  bookmarks=true,%
  bookmarksopen=true,%
  breaklinks=true,%
  colorlinks=true,%
  linkcolor=blue,anchorcolor=blue,%
  citecolor=blue,filecolor=blue,%
  menucolor=blue,pagecolor=blue,%
  urlcolor=blue]{hyperref}
\else
\usepackage[%
  breaklinks=true,%
  colorlinks=true,%
  linkcolor=blue,anchorcolor=blue,%
  citecolor=blue,filecolor=blue,%
  menucolor=blue,pagecolor=blue,%
  urlcolor=blue]{hyperref}
\fi
\usepackage{tabularx}
\usepackage{natbib}

\journal{Photosynthesis Research}

\makeatletter
\def\elsartstyle{%
    \def\normalsize{\@setfontsize\normalsize\@xiipt{14.5}}
    \def\small{\@setfontsize\small\@xipt{13.6}}
    \let\footnotesize=\small
    \def\large{\@setfontsize\large\@xivpt{18}}
    \def\Large{\@setfontsize\Large\@xviipt{22}}
    \skip\@mpfootins = 18\p@ \@plus 2\p@
    \normalsize
}
\@ifundefined{square}{}{}
\makeatother

\usepackage{listings}
\usepackage{xcolor}
\usepackage{float}

\lstdefinestyle{pythonstyle}{
    language=Python,
    basicstyle=\small\ttfamily,
    keywordstyle=\color{blue}\bfseries,
    commentstyle=\color{green!50!black},
    stringstyle=\color{red},
    frame=single,
    showstringspaces=false,
    numbers=left,
    numberstyle=\tiny\color{gray},
    breaklines=true
}

\begin{document}

\begin{frontmatter}
\title{PhoTorch: A robust and generalized biochemical photosynthesis model fitting package based on PyTorch}

\author{Tong Lei \corref{cor}\fnref{ucdps}}\ead{tlei@ucdavis.edu}
\author{Kyle T. Rizzo\fnref{ucdps}}
\author{Brian N. Bailey \fnref{ucdps}}

\cortext[cor]{Corresponding author}
\address[ucdps]{Department of Plant Sciences, University of California, Davis, Davis, CA USA}

\begin{abstract}

\doublespacing
Advancements in artificial intelligence (AI) have greatly benefited plant phenotyping and predictive modeling. However, unrealized opportunities exist in leveraging AI advancements in model parameter optimization for parameter fitting in complex biophysical models. This work developed novel software, PhoTorch, for fitting parameters of the Farquhar, von Caemmerer, and Berry (FvCB) biochemical photosynthesis model based the parameter optimization components of the popular AI framework PyTorch. The primary novelty of the software lies in its computational efficiency, robustness of parameter estimation, and flexibility in handling different types of response curves and sub-model functional forms. PhoTorch can fit both steady-state and non-steady-state gas exchange data with high efficiency and accuracy. Its flexibility allows for optional fitting of temperature and light response parameters, and can simultaneously fit light response curves and standard \(A/C_i\) curves. These features are not available within presently available \(A/C_i\) curve fitting packages. Results illustrated the robustness and efficiency of PhoTorch in fitting \(A/C_i\) curves with high variability and some level of artifacts and noise. PhoTorch is more than four times faster than benchmark software, which may be relevant when processing many non-steady-state \(A/C_i\) curves with hundreds of data points per curve. PhoTorch provides researchers from various fields with a reliable and efficient tool for analyzing photosynthetic data. The Python package is openly accessible from the repository: \url{https://github.com/GEMINI-Breeding/photorch}.

\end{abstract}

\end{frontmatter}

\doublespacing

\section{Introduction}\label{intro}

The Farquhar, von Caemmerer, and Berry (FvCB) model \citep{Farquhar80} is a widely used biochemically-based framework for describing the photosynthetic response of \(C_3\) plants to intercellular CO$_2$ concentration, with more recent variations including the impacts of light, temperature, and water status, among others. This model, along with its many modified versions, defines numerous physiologically-based parameters that can be used to describe the response of photosynthesis to environmental variables \citep{Farquhar80, harley92cotton, medlyn2002, sharkey2007fitting}. The extracted parameters are often used as quasi-traits in plant physiology, crop modeling, ecological modeling, and climate change research due to their mechanistic linkage to physiological processes \citep{martinez2018acclimation,kim2020use,li2021application,shin2021evaluation,xue2022quantitative}. 

The FvCB model parameters cannot be directly measured in isolation, but are instead typically inferred by fitting the FvCB model to photosynthetic response curves. This means that, in addition to measurement errors, errors can also be incurred from the model fitting procedure. Challenges inherent in fitting the FvCB to experimental response curves have been explored extensively in prior literature \citep[e.g.,][]{long2003gas,miao2009comparison,gu2010reliable,busch2024guide}. These challenges arise due to the fact that the model is over-parameterized, and is a switch-type equation with discontinuous transitions between states. This can create situations in which different combinations of model parameters result in similar fitting errors, and can make fitted values sensitive to initial parameter guesses. 

Currently available packages that specialize in fitting the FvCB model, which mitigate associated fitting challenges to some extent, include FitFarquharModel \citep{FitFarquharModel}, plantecophys \citep{plantecophys}, ``photosynthesis" \citep{stinziano2021principles}, and various Excel-based routines \citep{sharkey2007fitting,bellasio2016xcel}. Limitations of current models or packages include the inability to fit TPU parameters \citep{FitFarquharModel} and the inability to fit light and temperature response parameters simultaneously \citep{plantecophys, stinziano2021principles}. These packages do not provide fitting for a comprehensive set of CO\textsubscript{2}, light and temperature response parameters, and thus they may need to assume some parameters are conserved across different species or genotypes, even though parameter variability exists \citep{wieloch2023model,busch2024guide}. Excel-based models lack computational efficiency and flexibility, and can be highly sensitive to initial parameter guesses \citep{dubois2007optimizing}. These Excel-based methods also require users to specify the transition point from RuBP-saturated to RuBP-limited photosynthesis rate \citep{busch2024guide}. Additionally, these packages were intended for use with steady-state \(A/C_i\) response data, while more recent instruments allow for \(A/C_i\) curve estimation based on non-steady-state data \citep{stinziano2017rapid,taylor2019phenotyping,stinziano2019rapid,dynamicA,tejera2024dynamic}. As result, these packages can have long runtimes for dense non-steady-state \(A/C_i\) data, although this can be mitigated by downsampling the steady-state data to use fewer data points \citep{burnett2019one}, with the trade-off being the accuracy of parameter estimation. Therefore, developing new \(A/C_i\) curve fitting tools that can fit both steady-state and non-steady-state gas exchange response data, while also providing additional flexibility for fitting light and temperature response parameters, is of potential value to the community.

Advancements in artificial intelligence have also led to advancements in its related sub-disciplines, such as in efficient and robust parameter calibration within complex models. Artificial intelligence is effectively a parameter optimization problem, where the goal is to determine the values of thousands or millions of parameters that allow the model to optimally fit the training data. These parameter optimization routines have more generalized potential to improve parameter fitting in complex biophysical models.

Generalized frameworks providing a computational foundation for implementing AI models have emerged, such as the Python-based package PyTorch \citep{pytorchc}. As one of the most popular deep learning frameworks, PyTorch offers many highly optimized standard functions with high computational performance on both CPU and GPU hardware. Thus, encoding the FvCB model in PyTorch and treating it as a deep learning model, with the physiological parameters as adjustable weights, is both reasonable and feasible. The optimization of FvCB model parameters can utilize advanced features provided by PyTorch, such as automatic differentiation and flexible enforcement of constraints.

This work developed a new tool named PhoTorch based on PyTorch for determination of parameters of the biochemical FvCB photosynthesis model based on leaf-level gas exchange data. In addition to reliable parameter estimation, novel aspects of the tool include its flexibility in fitting to CO\textsubscript{2}, light, and temperature response data, inclusion of multiple light and temperature functions, and ability to handle noisy data or data with artifacts in some cases. It can handle both non-steady-state and steady-state \(A/C_i\) curves with high efficiency. These benefits are achieved through careful regulation of the optimization process and the full utilization of PyTorch's automatic differentiation and the computational efficiency of the Adam optimizer. The goal is to provide a user-friendly, open-source tool for accurately estimating photosynthetic parameters that improves users' ability to model photosynthetic responses under various temperature and light conditions.

\section{Materials and methods}\label{S:model}

\subsection{FvCB model with light and temperature response functions}\label{sect:fvcb}

The goal of the PhoTorch software is to estimate parameters in the FvCB model based on input leaf-level gas exchange data. This section describes the theory and forms of the FvCB used by PhoTorch, and the associated model parameters. The FvCB model describes the rate of photosynthesis as being limited by the rate of ribulose-1,5-bisphosphate carboxylase/oxygenase (Rubisco) carboxylation, the rate of ribulose-1,5-bisphosphate (RuBP) regeneration, or the rate of triose-phosphate utilization (TPU). The general model form used herein is based on that of \citet{harley92cotton} with two possible light response functions and two possible temperature response functions, and with the option of fitting the mesophyll conductance (\(g_m\)), as in many prior studies \citep{ethier2004}. The net rate of photosynthesis (\(A\)) on a per leaf area basis is calculated according to:

\begin{equation}\label{eq:fvcbmain}
A = \min(W_c, W_j, W_p)(1-\frac{\Gamma^*}{C_i}) - R_d,
\end{equation}

\noindent where \(R_d\) (\(\mu\)mol/m\(^2\)/s) is the dark respiration rate, \(\Gamma^*\) (\(\mu\)mol/mol) is the CO$_2$ compensation point when mitochondrial respiration is zero, and \(C_i\) (\(\mu\)mol/mol) is the intercellular CO$_2$ concentration. 

The rate of carboxylation limited by the amount, activation state, and kinetic properties of Rubisco (\(W_c\)) is given by:
\begin{equation}\label{eq:wc}
W_c = \frac{V_{cmax}C_i}{C_i + K_c \left(1 + \frac{O}{K_o}\right)},
\end{equation}

\noindent where \(V_{cmax}\) (\(\mu\)mol/m\(^2\)/s) is the maximum rate of carboxylation, \(K_c\) (\(\mu\)mol/mol) and \(K_o\) (mmol/mol) are the Michaelis-Menten constants for carboxylation and oxygenation, respectively, and \(O\) (mmol/mol) is the oxygen concentration.

The rate of carboxylation limited solely by the RuBP regeneration (\(W_j\)) is described by:
\begin{equation}\label{eq:wj}
W_j = \frac{JC_i}{4(C_i + 2\Gamma^*)},
\end{equation}

\noindent where \(J\)  (\(\mu\)mol/m\(^2\)/s) is the CO$_2$ saturated electron transport rate. Its response to the incoming photosynthetic photon flux density (PPFD) can be described using a rectangular hyperbola \citep{Farquhar80}:

\begin{equation}\label{eq:lr1}
J = \frac{\alpha Q J_{max}}{\alpha Q + J_{max}},
\end{equation}

\noindent where \(\alpha\) is the quantum yield of electron transport, \(Q\) (\(\mu\)mol/m\(^2\)/s) is the absorbed PPFD, and \(J_{max}\) is the maximum rate of electron transport. A non-rectangular hyperbolic function \citep{medlyn2002} is more widely used to describe the relationship between \(J\) and PPFD:

\begin{equation}\label{eq:lr2}
J = \frac{\alpha Q + J_{max} - \sqrt{(\alpha Q + J_{max})^2 - 4 \theta \alpha Q J_{max}}}{2 \theta},
\end{equation}

\noindent where \(\theta\) is the curvature parameter. Note that as $\theta$ approaches zero, Eqs. \ref{eq:lr1} and \ref{eq:lr2} become equivalent.

In the model fitting code, users have the option to use Eq. \ref{eq:lr1} (rectangular hyperbola) or Eq. \ref{eq:lr2} (non-rectangular hyperbola), and are referenced respectively as light response type 1 and 2. If the light response type is set to 0, \(J\) is always equal to \(J_{max}\).

The rate of carboxylation limited by inorganic phosphate (\(W_p\)) is given by:
\begin{equation}\label{eq:wp}
W_p = 3\mathrm{TPU}\frac{C_i}{C_i-(1+3\alpha_g)\Gamma^*},
\end{equation}
\noindent where \(\mathrm{TPU}\) (\(\mu\)mol/m\(^2\)/s) is the rate of triose phosphate utilization, and \(\alpha_g\) is the stoichiometric ratio of orthophosphate (P\(_i\)) consumption in oxygenation \citep{voncae2000,ellsworth2015phosphorus}. 

If the user includes fitting of mesophyll conductance \(g_m\), then the above-mentioned \(C_i\) is replaced with chloroplastic CO$_2$ concentration \(C_c\) according to:

\begin{equation}
C_c= C_i - A/g_m.
\end{equation}

The temperature responses of four main parameters \(V_{cmax}\), \(J_{max}\), \(\mathrm{TPU}\), and \(R_d\) of the FvCB model can be represented according to the Arrhenius function \citep{medlyn2002}:

\begin{equation}\label{eq:tpr1}
k = k_{25} \exp{\left[\frac{\Delta{H_a}}{R}\left(\frac{1}{298}-\frac{1}{T_{leaf}}\right)\right]},
\end{equation}

\noindent which results in monotonic increase in the parameter ($k$ being one of \(V_{cmax}\), \(J_{max}\), \(\mathrm{TPU}\), or \(R_d\), and \(k_{25}\) is the value of the corresponding parameter at 25\(^\circ\)C) with leaf temperature $T_{leaf}$ (Kelvin). A more flexible temperature response function can also be used that can represent a temperature optimum:

\begin{equation}\label{eq:tpr2}
k = k_{25} \exp\left[\frac{\Delta H_a}{R} \left(\frac{1}{298}-\frac{1}{T_{leaf}}\right)\right]  \frac{f\left(298\right)}{f\left(T_{leaf}\right)},
\end{equation}

\begin{equation}
    f(T) = 1+\exp \left[\frac{\Delta H_d}{R}\left(\frac{1}{T_{opt}}-\frac{1}{T} \right)-\ln \left(\frac{\Delta H_d}{\Delta H_a}-1 \right) \right]
\end{equation}

\noindent where \(\Delta{H_a}\) is the activation energy (kJ/mol), \(\Delta{H_d}\) is the deactivation energy (kJ/mol), \(T_{opt}\) is the temperature in Kelvin at which \(k\) is optimal, and \(R\) is the universal gas constant (0.008314 KJ/mol/K). 

\(K_c\), \(K_o\), and \(\Gamma^*\) can be directly calculated based on Eq. \ref{eq:tpr1} using their corresponding \(k_{25}\) and \(\Delta{H_a}\) values reported by \citep{bernacchi2001}, or they can be directly fitted using PhoTorch. It should be noted that the parameter \(c\) used in the temperature response function of \citet{bernacchi2001} is a scaling constant, which is equal to \(\frac{\Delta H_a}{298R}+\ln{k_{25}}\), and thus can be substituted for and omitted as a parameter. In the code, the temperature response types 1 and 2 correspond to Eq. \ref{eq:tpr1} and \ref{eq:tpr2}, respectively. If the temperature response type is set to 0,  \(V_{cmax}\), \(J_{max}\), \(\mathrm{TPU}\), and \(R_d\) will be directly fitted, and no temperature response will be applied. Table \ref{tab:paraml} and \ref{tab:paramd} lists the above-mentioned fitted and fixed parameters, respectively. All of these parameters, except \(\alpha_g\), can be replaced by user specified values.

\begin{table}
\centering
\caption{Parameters in the FvCB model, and conditions under which the parameters are fitted or specified as constants by the user in PhoTorch.}
\label{tab:paraml}
\begin{tabular}{>{\raggedright\arraybackslash}p{2.2cm} >{\raggedright\arraybackslash}p{6.2cm}>{\raggedright\arraybackslash}p{2cm}>{\raggedright\arraybackslash}p{3.7cm}}
  \hline
  Parameter & Condition when fitted & Default value & Reference\\
  \hline
  \(V_{cmax25}\) (\(\mu\)mol/m\(^2\)/s) & Always & 100 & - \\
  \(J_{max25}\) (\(\mu\)mol/m\(^2\)/s)  & Always & 200 & - \\
  \(\mathrm{TPU_{25}}\) (\(\mu\)mol/m\(^2\)/s)  & $^a$Always & 25 & - \\
  \(R_{d25}\) (\(\mu\)mol/m\(^2\)/s) & Always & 1.5 & - \\
  \(K_{c25} \) (\(\mu\)mol/mol) & Specified by user & 404.9 & \citep{bernacchi2001}\\
  \(K_{o25} \) (mmol/mol) & Specified by user & 278.4 & \citep{bernacchi2001}\\
  \(\Gamma_{25}^* \) (\(\mu\)mol/mol) & Specified by user & 42.75 & \citep{bernacchi2001}\\
  \(\Delta{H_{a,V_{cmax}}}\) (kJ/mol)& Temperature response type 1 or 2 used & 65.33 & \citep{sharkey2007fitting}\\
  \(\Delta{H_{a,J_{max}}}\) (kJ/mol)& Temperature response type 1 or 2 used & 43.9 & \citep{sharkey2007fitting}\\
  \(\Delta{H_{a,\mathrm{TPU}}}\) (kJ/mol)& Temperature response type 1 or 2 used & 53.1 & \citep{harley92cotton}\\
  \(T_{opt,V_{cmax}}\) (K)& Temperature response type 2 used & $311$ & \citep{medlyn2002}\\
  \(T_{opt,J_{max}}\) (K)&Temperature response type 2 used & $311$ & \citep{medlyn2002}\\
  \(T_{opt,\mathrm{TPU}}\) (K)& Temperature response type 2 used & $^b306$ & \citep{harley92cotton}\\
  \(\alpha_g\) & Always & 0 & - \\
  \(g_m\) (mol/m\(^2\)/s) & Specified by user & 10 & - \\
  \(\alpha\) & Light response type 1 or 2 used & 0.5 & - \\
  \(\theta\) & Light response type 2 used & 0.7 & - \\
  \hline
\end{tabular}
\parbox{\textwidth}{
\small {$^a$: If only the light response curve is specified by the user, $TPU_{25}$ will not be fitted. \par
$^b$: Derived from entropy $\Delta{S}$ of 0.65 \citep{harley92cotton} based on the equation: \(\Delta{S}=\Delta{H_d}/T_{opt}+R\log{\frac{\Delta{H_a}}{\Delta{H_d}-\Delta{H_a}}}\)}}
\end{table}

\begin{table}
\centering
\caption{Additional fixed temperature response parameters in the FvCB model.}
\label{tab:paramd}
\begin{tabular}{>{\raggedright\arraybackslash}p{3.2cm} >{\raggedright\arraybackslash}p{2.1cm}>{\raggedright\arraybackslash}p{3.8cm}}
  \hline
  Parameter & Default value & Reference\\
  \hline
  \(\Delta{H_{a,R_d}}\) (kJ/mol)& 46.39 & \citep{sharkey2007fitting}\\
  \(\Delta{H_{a,K_c}}\) (kJ/mol)& 79.43 & \citep{bernacchi2001}\\    
  \(\Delta{H_{a,K_o}}\) (kJ/mol)& 36.38 & \citep{bernacchi2001}\\
  \(\Delta{H_{a,\Gamma^*}}\) (kJ/mol)& 37.83 & \citep{bernacchi2001}\\
  \(\Delta{H_{d,V_{cmax}}}\) (kJ/mol)& 200 & \citep{medlyn2002}\\
  \(\Delta{H_{d,J_{max}}}\) (kJ/mol)& 200 & \citep{medlyn2002}\\
  \(\Delta{H_{d,\mathrm{TPU}}}\) & 201.8 & \citep{sharkey2007fitting}\\
  \hline
\end{tabular}
\end{table}

\subsection{Supported input gas exchange data}\label{S:input}

PhoTorch accepts various types of steady-state and non-steady-state leaf-level gas exchange data. Photosynthetic response curves are typically generated by placing all or a portion of a leaf within a cuvette with specified environmental conditions (e.g., ambient CO\textsubscript{2} concentration, light flux, air/leaf temperature), and changing one environmental variable at a time across a specified range. Steady-state gas exchange data refers to measurements where the cuvette environment and photosynthesis measurement remains ``stable" at each measured data point, which results in a typical total measurement time lasting 30-60 minutes and a lower curve data density (usually around 10 points). In contrast, more recently developed non-steady-state gas exchange measurements involve a continuous ramp of the response variable during measurement, resulting in higher measurement data density (over 100 points) and faster measurement times (typically around 10 minutes). Two of the most common non-steady-state measurements techniques are RACiR \citep{stinziano2017rapid} and the dynamic assimilation technique \citep[DAT;][]{dynamicA}.

PhoTorch can accept two types of photosynthetic response curves: response to variable ambient CO\textsubscript{2} concentration (hereafter referred to as ``\(A/C_i\) curves", and response to variable light flux (hereafter referred to as ``\(A/Q\)" curves. In either case, all variables other than the response variable should be held constant (some fluctuation is tolerable), and the response variable is varied in incremental steps (steady-state) or continuously (non-steady-state) while logging the corresponding net CO\textsubscript{2} flux from the leaf and corresponding environmental conditions. 

To fit temperature response parameters, multiple \(A/C_i\) or \(A/Q\) curves with different leaf temperatures should be included in the input dataset. In order to reliably fit the temperature response curve with an optimum (temperature response type 2; Eq. \ref{eq:tpr2}), temperatures above and below the photosynthetic temperature optimum should be included.

\subsection{Input data format}\label{S:input_format}

Table \ref{tab:csvtemp} shows an example of an input ``CSV" file, which is the format required by PhoTorch. The input files can be easily created in Microsoft Excel, for example, and exported in CSV format. The data from each response curve is placed in the input file, with rows corresponding to each CO\textsubscript{2} measurement point and associated environmental conditions. Multiple response curves are simply placed in the file one after the other, with each curve given an arbitrary curve ID value to distinguish them. The input data file must have a header row with columns titles of ``CurveID", ``FittingGroup", ``Ci", and ``A" at a minimum. The ``CurveID" should be a unique integer for each response curve. ``FittingGroup" is also an integer that denotes grouping of multiple curves for parameter fitting. Curves with the same ``FittingGroup" value will be aggregated to produce only one overall fitted parameter set for all parameters outside of the four main parameters \(V_{cmax25}\), \(J_{max25}\), \(TPU_{25}\), and \(R_{d25}\). For example, all curves with FittingGroup=1 would be aggregated to produce a single fitted $\Delta H_{a,Vcmax}$, $\Delta H_{a,Jmax}$, etc., but each curve would still have its own unique \(V_{cmax25}\), \(J_{max25}\), \(TPU_{25}\), and \(R_{d25}\) values. If the ``onefit" option is set to ``True", curves in the same group will also share these four main parameters. As defined above, ``\(C_i\)" is the intercellular CO\textsubscript{2} concentration in units of $mu$mol/mol, and ``\(A\)" is the net CO\textsubscript{2} flux in $\mu$mol/m$^2$/s.

Additionally, columns can be provided to specify the PPFD in units of $\mu$mol/m$^2$/s (column label of ``Qin"), or the leaf surface temperature in units of Celsius (column label of ``Tleaf").  If ``Qin" and ``Tleaf" are not provided, default values of 2000 and 25 will be used for these two parameters. Other columns may be present in the file, but they will be ignored by the software.

\begin{table}[h!]
\centering
\caption{Example of an input ``CSV" data file for two \(A/C_i\) curves collected at two different temperatures. Ci is the intercellular CO$_2$ concentration (\(\mu\)mol/mol), A is the net rate of photosynthesis (\(\mu\)mol/m\(^2\)/s), Qin is the absorbed PPFD (\(\mu\)mol/m\(^2\)/s), and Tleaf is the leaf temperature (Kelvin).  Note that data values are not representative of real \(A/C_i\) curves, as they has been shortened for brevity.}
\begin{tabular}{|c|c|c|c|c|c|}
\hline
CurveID & FittingGroup & Ci & A & Qin & Tleaf \\
\hline
1 & 1 & 200 & 20 & 2000 & 25 \\
1 & 1 & 800 & 30 & 2000 & 25 \\
1 & 1 & 1500 & 40 & 2000 & 25 \\
2 & 1 & 200 & 25 & 2000 & 30 \\
2 & 1 & 800 & 35 & 2000 & 30 \\
2 & 1 & 1500 & 55 & 2000 & 30 \\
\hline
\end{tabular}
\label{tab:csvtemp}
\end{table}

\subsection{Example Python code for parameter fitting}

Users can directly download or clone all files from the GitHub repository to their own working directory (replace the example ``CSV" file with users' own ``CSV" data file). Once the input data file has been prepared, model fitting can be performed by running a simple Python script based on the `fitaci' package (\url{https://www.github.com/GEMINI-Breeding/photorch}).

Listing \ref{lst:python_example} provides a minimal Python script example for performing model fitting. This essentially involves loading and initializing the data, initializing the optimizer, and performing the fitting. Users should consult the software documentation for information on how to utilize more advanced features of PhoTorch.

\begin{lstlisting}[style=pythonstyle, caption={Minimal example Python code to fit FvCB parameters based on an input CSV file. The code assumes you have created a properly formatted CSV file containing the photosynthesis gas exchange data (see Sect. \ref{S:input_format} and Table \ref{tab:csvtemp}), and placed it in the current working directory or Python path. Additionally, dependent packages `torch', `numpy', `scipy' and `pandas' must be installed (e.g., using pip - `\texttt{pip install torch}'). Resulting parameters will be stored in the variable `fvcbm'.}, captionpos=b, label={lst:python_example}, float=p]
import fitaci
import pandas as pd
import torch

# Read in the CSV data.
dftest = pd.read_csv("mydata.csv")

# Initialize and pre-process the data
lcd = fitaci.initD.initLicordata(dftest, preprocess=True)

# Initialize the parameter optimizer
fvcbm = fitaci.initM.FvCB(lcd, LightResp_type = 0, TempResp_type = 2, onefit = False)

# Run the parameter fitting
fitresult = fitaci.run(fvcbm)

# Store result into a structure called "fvcbm" that contains all the fitted parameters
fvcbm = fitresult.model
\end{lstlisting}

\subsection{Parameter fitting algorithm details}

\subsubsection{Automated \(A/C_i\) curve pre-processing}\label{S:pe}

\begin{figure}
\begin{center}
\includegraphics[width=8cm]{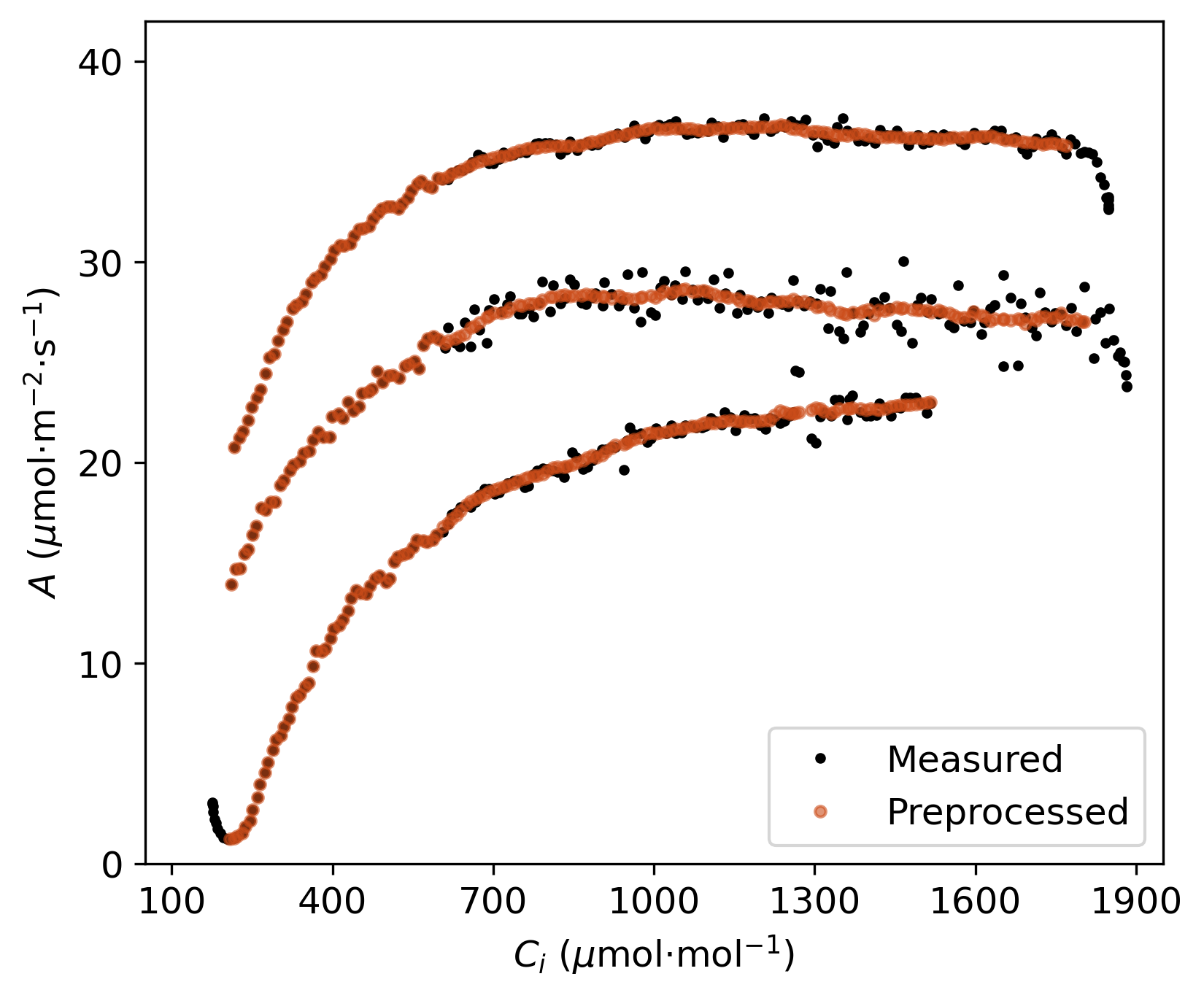}
\end{center}
\caption{Example illustration of optional automated pre-processing of measured $A/C_i$ curves to smooth and remove artifacts at the beginning and end of the curves. The black and red points represent measured raw and pre-processed sample points, respectively. }
\label{fig:preprocessing}
\end{figure}

Noise and artifacts in photosynthetic response curves derived from gas exchange measurements, especially for non-steady-state measurements, can greatly affect FvCB model fitted parameters. Therefore, an optional pre-processing procedure is integrated to automatically remove noisy data points before model fitting when possible (Fig. \ref{fig:preprocessing}). For \(C_i\) larger than 600 \(\mu\)mol/mol, a first order Savitzky-Golay smoothing filter \citep{sgsmooth} with a default window length of 10 is applied. Additionally, data points at the ends of the curves that exhibit excessive upward or downward variation are removed. Excessive upward variation was defined as a difference between two consecutive $A$ data points greater than 0.06 \(\mu\)mol/m\(^2\)/s, and excessive downward variation was defined as a difference less than -0.06 \(\mu\)mol/m\(^2\)/s. If the number of data points is less than an adjustable threshold (the default threshold is 3 times the window length), the smoothing will not be applied, making it generalized to steady-state gas exchange data. Data points in which \(C_i\) is less than the \(C_i\) value at the point with the minimum \(A\), and where \(A\) is less than the \(A\) value at the point with the minimum \(C_i\) are also removed. These settings can be adjusted by users and are automatically skipped for light response curves.

\subsubsection{PyTorch parameter optimizer (Adam)}

As a popular deep learning architecture, PyTorch \citep{pytorchc} provides many important features that are beneficial for fitting photosynthesis curves. It can automatically compute gradients, which refers to the derivative of the loss function with respect to the parameters, of a customized model through a process called automatic differentiation. The loss function (i.e., error function to be minimized during fitting) can also be flexibly customized. Furthermore, some built-in activation functions can help constrain the range of FvCB parameters. For example, the sigmoid function $\sigma(x) = \frac{1}{1 + e^{-x}}$ can ensure \(\alpha_g\) remains within the range of 0 to 1.
  
The Adaptive Moment Estimation (Adam) optimizer \citep{Adam} was the chosen algorithm for fitting the FvCB model bases on photosythesis response curves. Although PyTorch provides other optimizers such as stochastic gradient descent (SGD), these require more tuning of hyperparameters and are more suitable for conventional machine learning models. Adam uses first-order gradient-based optimization of stochastic objective functions, based on adaptive estimates of lower-order moments. Given an initial guess for a model parameter \(k\) (e.g., $V_{cmax25}$), the value is updated iteratively according to:

\begin{equation}\label{eq:adam}
k_t = k_{t-1} - l \frac{\hat{m}_t}{\sqrt{\hat{v}_t} + \epsilon},
\end{equation}

\noindent where \(t\) is the updating iteration, \(l\) is the learning rate which determines how aggressively the parameter estimate is updated on each iteration (the default is set to 0.08 in PhoTorch), \(\hat{m}_t\) represents the biased first moment estimate, which determines the direction of optimization and is calculated as exponential moving averages of the gradients, \(\hat{v}_t\) denotes the biased second raw moment estimate, which scales down the step size in regions where gradients might cause instability and is calculated as the squared gradients, \(\epsilon\) is a hyperparameter which is set to \(10^{-8}\). Adam can automatically adjust the learning rate \(l\) for each FvCB parameter individually based on estimates of the first and second moments of the gradients. This feature makes Adam very effective in FvCB model fitting, as the parameters of the FvCB model greatly vary in terms of their sensitivity, units, and magnitude. Additionally, Adam requires little tuning of hyperparameters and has high optimization efficiency. The default values of two hyperparameters (learning rate 0.08 and maximum iteration 20,000) work for most \(A/C_i\) fitting cases.

\subsubsection{Loss function}\label{S:lf}

The loss function is the equation to be minimized by the optimizer, and quantifies the level of agreement between the measured photosynthesis data points and the FvCB model prediction given the current parameter set. The design of the loss function is crucial for fitting the FvCB model in order to obtain a reliable and accurate fit for arbitrary input data sets. The basic loss function is the mean squared error (\(\mathrm{MSE}\)) between the measured \(A_i\) and predicted \(\hat{A_i}\) net photosynthesis rates for a given parameter set:

\begin{equation}\label{eq:mse}
\mathrm{MSE} = \frac{1}{n} \sum\limits_{k=1}^n (A_k - \hat{A}_k)^2,
\end{equation}

\noindent where \( n \) is the total number of measured data points. 

Equation \ref{eq:mse} forms the basis of the loss function, but additional penalty terms are added to constrain the model fit in order to increase robustness. Penalty terms are essentially additional errors added to the loss to enforce specific properties in the output curves (e.g., parameter constraints). With increasing $C_i$ we assume that the assimilation will progress from being Rubisco limited, to be being limited by RuBP regeneration, and then finally to being limited by triose phosphate utilization. We assume all response curves contain Rubisco, RuBP, and TPU limitation states, and the order of these three states is fixed. According to Eq. \ref{eq:wc}, \ref{eq:wj} and \ref{eq:wp}, as long as leaf temperature and incident light remain nearly constant (though some fluctuation may exist), curves of the Rubisco-limited assimilation rate \(A_c(C_i)\) and the RuBP-limited assimilation rate \(A_j(C_i)\) remain nearly monotonically increasing, and the curve of the TPU-limited assimilation rate \(A_p(C_i)\) remains nearly constant (\(\alpha_g=0\)) or nearly monotonically decreasing (\(\alpha_g>0\)). Thus, the intersection point between \(A_c(C_i)\) and \(A_j(C_i)\) should always be lower than the point on \(A_p(C_i)\) with the same \(C_i\), which can be expressed as:

\begin{equation}
C_{i,jc} = \arg \min_{k} | A_j(C_{i,k}) - A_c(C_{i,k}) |
\end{equation}

\noindent where \(A_c\) and \(A_j\) curves are closest to each other at \(C_{i,jc}\), and \(|\cdot|\) denotes the absolute values. Then, the penalty can be given as:

\begin{equation}\label{eq:pentalty_1}
\text{Penalty}_{cj>p} = \max[0,\quad\max[A_j(C_{i,jc}), A_c(C_{i,jc})]-A_p(C_{i,jc})].
\end{equation}

The above penalty targets the closest points between the curves rather than the actual curve intersection points. The reason is that the \(A/C_i\) curves are based on discrete data, so in most cases an exact intersection point does not exist. Therefore, penalties that force \(A_c(C_i)\) and \(A_j(C_i)\) to intersect each other should be added:

\begin{equation}\label{eq:pentalty_2}
\text{Penalty}_{c>j} = \max\left[\beta - \sum \max\left[A_c(C_i) - A_j(C_i), \quad0\right],\quad0\right],
\end{equation}

\begin{equation}\label{eq:pentalty_3}
\text{Penalty}_{c<j} = \max\left[\beta - \sum \max\left[A_j(C_i) - A_c(C_i), \quad0\right],\quad0\right],
\end{equation}

\noindent where \(\beta\) is a constant specified by the user (default is 8). The \(\text{Penalty}_{c>j}\) and \(\text{Penalty}_{c<j}\) terms penalize the situation where \(A_c(C_i)\) is supposed to be greater than \(A_j(C_i)\) if the sum less than \(\beta\) and where \(A_j(C_i)\) is supposed to be greater than \(A_c(C_i)\) if the sum less than \(\beta\), respectively. In other words, these penalties are supposed to be zero only if \(A_c(C_i)\) and \(A_j(C_i)\) intersect with each other sufficiently. 

An additional constraint is added to ensure that a transition to TPU limitation occurs within the measured response curve, otherwise the gradient with respect to \(TPU_{25}\) will vanish. Thus, a penalty is added when the last point (i.e., highest measured $C_i$) of \(A_p\) is greater than that of \(A_j\): 

\begin{equation}\label{eq:pentalty_4}
\text{Penalty}_{j<p} = \max[0, A_p(C_{i,last})]-A_j(C_{i,last})].
\end{equation}

\noindent This penalty forces \( A_p(C_i)\) to move down until its last point is lower than that of \(A_j(C_i)\). As \(A_j(C_i)\) is nearly monotonically increasing and \(A_p(C_i)\) is flat or monotonically decreasing, the penalty has little effect on the final reconstruction even if the \(A/C_i\) curve has no TPU limitation stage, ensuring that the parameter (\(TPU\) or \(TPU_{25}\)) always has gradient. After fitting, the difference $A_p(C_{i,last})-A_j(C_{i,last})$ can be used to identify whether the target \(A/C_i\) curve has a TPU limitation stage. This penalty can also be turned off if the user is fitting the light response curve. 

As previously reported, \(V_{cmax}\) and \(J_{max}\) are highly correlated \citep{wullschleger1993biochemical,medlyn2002, fan2011determination}. Thus, an optional penalty of a low correlation coefficient \(r\) between \(V_{cmax}\) and \(J_{max}\) was added to prevent the occurrence of anomalous values in some cases. Users can choose to turn this penalty on or off, and this penalty is only used when at least 7 curves are being fitted simultaneously. The correlation coefficient \(r\) is defined as:

\begin{equation}\label{eq:r}
r(x,y) =  \frac{\sum\limits_i (x_i - \overline{x})(y_i - \overline{y})}{\sqrt{\sum\limits_i (x_i - \overline{x})^2 \sum\limits_i (y_i - \overline{y})^2}},
\end{equation}
\noindent where \(\overline{x}\) and \( \overline{y}\) are the mean values of all data points in \(x\) and and \(y\) respectively. Then the penalty can be defined as:

\begin{equation}\label{eq:penalty_5}
\text{Penalty}_{r} = \max[0, 0.7-r(V_{cmax}, J_{max})].
\end{equation}

According to Eq. \ref{eq:penalty_5}, when \(r(V_{cmax}, J_{max})\) is less than 0.7, \(\text{Penalty}_{r}\) will be positive; otherwise, it is 0, which means no penalty. As illustrated by previous work \citep[e.g.,][]{netphoto}, a perfect correlation between $V_{cmax}$ and $J_{max}$ is not expected. Accordingly, this penalty is not particularly strict and is only meant to minimize the presence of outlier values.

For small non-negative parameters such \(\Delta{H_a}\) or \(R_{d25}\), penalties are also added to prevent negative values:

\begin{equation}\label{eq:penalty_6}
\text{Penalty}_{k<0} = \max[0, -k].
\end{equation}

The overall loss to be optimized is the sum of above mentioned \(MSE\) (Eq. \ref{eq:mse} and penalties (Eqs. \ref{eq:pentalty_1}, \ref{eq:pentalty_2}, \ref{eq:pentalty_3}, \ref{eq:pentalty_4}, \ref{eq:penalty_5},
\ref{eq:penalty_6}). 

\subsection{Field gas exchange data for method evaluation}

Field gas exchange data was collected across variable CO\textsubscript{2}, light, and temperature regimes in order to evaluate the proposed model calibration methodology. Measurements were taken within a common/tepary bean ({\em Phaseolus vulgaris} and {\em Phaseolus acutifolius}) and cowpea ({\em Vigna unguiculata}) breeding trial using an LI-6800 Portable Photosynthesis System ({LI-COR Biosciences}, Lincoln, Nebraska, USA) from June to August, 2022. The field was located in Davis, CA, United States. The breeding trials consisted of diversity panels of cowpea and bean genotypes, from which 5 plots of each were selected that represented high genetic diversity in terms of flowering time, yield, and growth structure, which also carried many traits relevant to abiotic and biotic stress resistance. Table \ref{tab:data} illustrates the details of the collected gas exchange data, and their pedigrees are listed in Table \ref{tab:pedigrees} of \ref{App:pedigrees}. Measurements were made on fully sunlit and fully expanded, healthy leaves between the hours of 9:00 and 12:00. \(A/C_i\) curves were run in non-steady-state mode using the LI-6800's Dynamic Assimilation Technique (DAT). Light response curves were obtained through separate response runs based on steady-state measurements at each light level. For the \(A/C_i\) curves, the reference relative humidity was set to 50\%, the incident light flux was set to 2000 \(\mu\)mol/m\(^2\)/s, the CO\textsubscript{2} ramp rate was 300 \(\mu\)mol/mol/min for the DAT, the fan speed was 10,000 rpm, and the starting and ending CO\textsubscript{2} concentration setpoints were 200 and 1800 \(\mu\)mol/mol, respectively. The leaf temperature setpoint was varied to be approximately 25, 30, and 40\degree{C}, and unique leaves were sampled for each \(A/C_i\) curve at the different temperatures. On each sampling day, all genotypes were sampled at all temperatures, with each curve being on a different leaf, and all measurements across multiple sampling days for a given genotype were aggregated to extract temperature parameters. All genotypes were sampled at a given temperature before moving on to the next higher temperature. For the light response curve, relative humidity was set to 60\%, leaf temperature was fixed to 25\degree{C}, fan speed was 3,000 rpm, incident PPFD values were set to 0, 150, 400, 800, 1200, and 2000 \(\mu\)mol/m\(^2\)/s. 

Oak ({\em Quercus coccinea} Münchh) \(A/C_i\) curves measured in steady-state from an openly accessible dataset \citep{burnett2021seasonal} were also used to evaluate PhoTorch's fitting ability. It contains 6 \(A/C_i\) curves measured using an LI-6800 and 35 using a LICOR LI-6400. Incident PPFD values were set to 1500 \(\mu\)mol/m\(^2\)/s, and leaf temperature ranged from 22 to 28\degree{C}. The dataset can be downloaded from Ecological Spectral Information System \citep{EcoSIS}.

\begin{table}
\centering
\caption{Overview of cowpea and bean gas exchange datasets used for method evaluation.}
\begin{tabular}{>{\raggedright\arraybackslash}p{3.7cm} >{\raggedright\arraybackslash}p{2.5cm}>{\raggedright\arraybackslash}p{2.0cm}>{\raggedright\arraybackslash}p{2.5cm}}
  \hline
  Species & Line ID & Number of curves & Number of total data points \\
  \hline
  Cowpea & $^a$MAGIC015 & 11 & 2027\\
  Cowpea & MAGIC179 & 11 & 1857\\
  Cowpea & MAGIC043 & 11 & 1836\\
  Cowpea & MAGIC043 & $^b$1 & 6\\
  Cowpea & MAGIC143 & 11 & 1857\\
  Cowpea & MAGIC184 & 11 & 1833\\
  Tepary Bean & $^c$TARS-Tep 23 & 13 & 2125\\
  Interspecific Common/Tepary Bean & $^d$NE-80-21-46 & 11 & 1902\\
  Interspecific Common/Tepary Bean & NE-80-21-2 & 12 & 2054\\
  Interspecific Common/Tepary Bean & NE-80-21-233 & 12 & 2028\\
  Interspecific Common/Tepary Bean & $^e$INB 841 & 12 & 2045\\
  \hline
\end{tabular}
\parbox{\textwidth}{
\small {$^a$: These ``MAGICXXX" genotypes are from the cowpea MAGIC population derived from eight parents \citep{huynh2018multi}. \par
$^b$: Steady-state light response curve. \par
$^c$: TARS-Tep 23 is a tepary line released by \citet{porch2022release}. \par
$^d$: The three ``NE-80-21-XX" common bean/tepary bean interspecific lines with prefixes of NE are common bean/tepary bean interspecific lines that were developed from crosses described in \citet{barrera2022large}. \par
$^e$: INB 841 is a common bean/tepary bean interspecific line that was developed from 
five cycles of congruity backcrossing of tepary with ICA Pijao \citep{mejia1994interspecific}.}}
\label{tab:data}
\end{table}

\subsection{Evaluation of method performance}

The coefficient of determination \(R^2\) and the root mean square error (\(\mathrm{RMSE}\)) were used as evaluation metrics to quantify the goodness of model fit.
The \(R^2\) for each curve fit is given as:

\begin{equation}\label{eq:r2}
R^2 = 1 - \frac{\sum\limits_i (A_i - \hat{A}_i)^2}{\sum\limits_i (A_i - \bar{A})^2},
\end{equation}
\noindent where \(\bar{A}\) is the mean value of all data points in \(A\), and \(\hat{A}\) is the fitted \(A\). The \(\mathrm{RMSE}\) is simply the square root of \(\mathrm{MSE}\) in Eq. \ref{eq:mse}:

\begin{equation}\label{eq:rmse}
\mathrm{RMSE} =\sqrt{ \frac{1}{n} \sum\limits_i (A_i - \hat{A}_i)^2}.
\end{equation}

For curves in one genotype, the model will fit their own four main parameters \(V_{cmax25}\), \(J_{max25}\), \(\mathrm{TPU_{25}}\), and \(R_{d25}\). For the light and temperature parameters, all curves for a given genotype will share the same set of parameters.

The plantecophys package \citep{plantecophys} was used as a benchmark for comparison. As plantecophys does not have the option for temperature response fitting, the temperature response option of PhoTorch was set to 0 when used for these comparisons. \(\alpha\), \(\Delta{H_{a,V_{cmax}}}\) and \(\Delta{H_{a,J_{max}}}\) of PhoTorch were set to 0.24, 58.5 and 29.68, respectively, for comparison, in accordance with default settings of plantecophys. No pre-processing procedure was applied to the original data. It should be noted that the error metric output from plantecophys is actually the `root sum squared error', although they call it `root mean squared error' (\(\mathrm{RMSE}\)). Therefore, plantecophys fitting results were used to calculate \(RMSE\) as defined in Eq. \ref{eq:rmse} for consistency in comparison with PhoTorch.

PhoTorch was also tested by fitting to synthetic \(A/C_i\) curves generated using the FvCB model, with known parameters obtained by randomly scaling those extracted by PhoTorch with a positive $R_d$ constraint (PhoTorch ($\pm R_d$)) in Table \ref{tab:compare} by 90\% to 110\%. Different levels of Gaussian noise were also added to the synthetic \(A/C_i\) curves. The number of added Gaussian noise data points is the same as the number of data points in a target \(A/C_i\) curve. The noise data points are independently sampled with a mean of zero and a varying standard deviation using the function \texttt{torch.randn()}.

\section{Results}

\subsection{Non-steady-state \(A/C_i\) curves at variable temperatures}

\begin{figure}
\centering
\includegraphics[width=10cm]{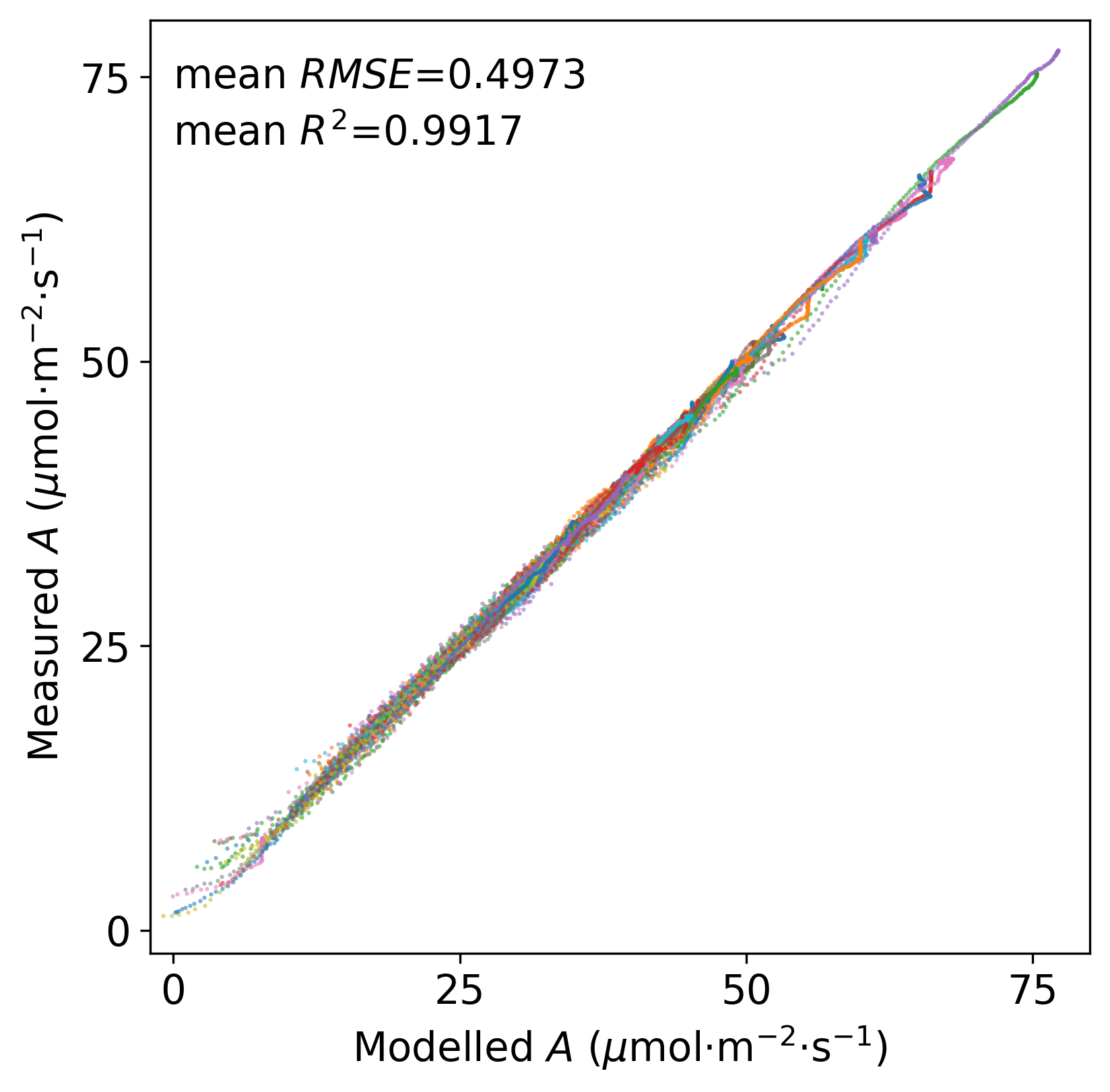}
\caption{Results of \(A/C_i\) fitting based on bean and cowpea field gas exchange data, excluding the light response curve. The horizontal and vertical coordinate of each point respectively corresponds to the modelled and measured \(A\) values at each \(C_i\) value. The learning rate and maximum iterations were set to 0.08 and 20,000 for all fittings. }
\label{fig:aafit}
\end{figure}

Figure \ref{fig:aafit} shows the result of \(A/C_i\) curve fitting for the bean and cowpea datasets at constant light and based on an Arrhenius temperature response (light response type of 0 and temperature response type of 1). The mean \(\mathrm{RMSE}\) and  \(R^2\) are 0.4973 (\(\mu\)mol/m\(^2\)/s) and 0.9917, respectively, using the same optimizer settings across all curves, indicating a good fit. Figure \ref{fig:acifit} presents more detailed fitting results. The mean \(\mathrm{RMSE}\) for fittings across all curves is within the range of 0.38-0.6. The resulting fit was nearly perfect with all \(R^2\) above 0.99 except the `Tep 23' curve, which still has a high \(R^2\) of 0.972 (Fig. \ref{fig:acifit}). These results demonstrate the robust performance of PhoTorch.

\begin{figure}
\centering
\includegraphics[width=8cm]{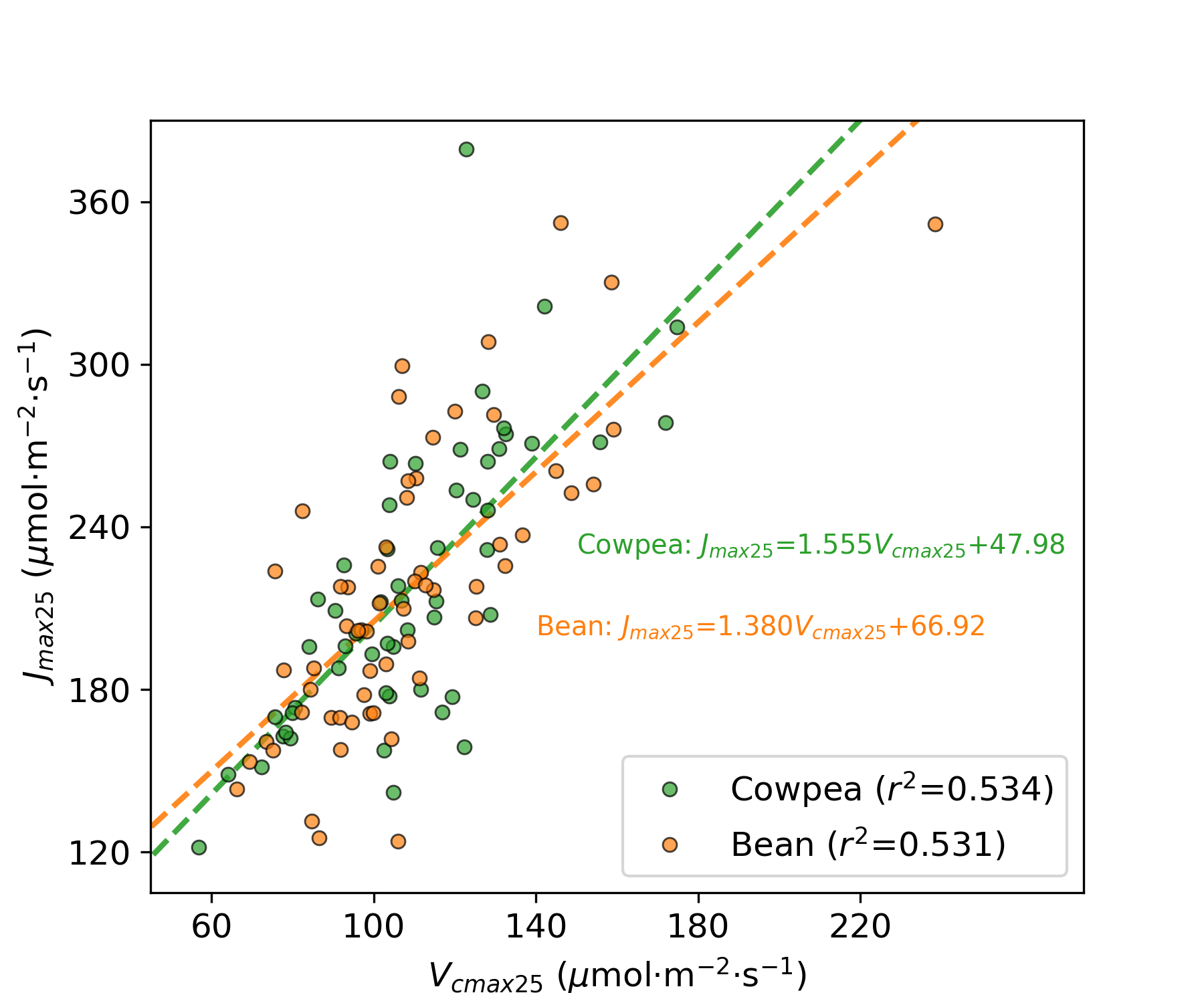}
\caption{Fitted \(V_{cmax25}\) versus \(J_{max25}\) for each \(A/C_i\) curve in the cowpea and bean datasets. Dashed lines give the best linear fit for each species.}
\label{fig:v_j}
\end{figure}

Resulting fitted parameters showed an approximately linear relationship between \(V_{cmax25}\) and \(J_{max25}\) (Fig. \ref{fig:v_j}), which is in agreement with prior observations \citep{wullschleger1993biochemical,medlyn2002,netphoto,kumarathunge2019acclimation} and supported by theoretical arguments of optimal allocation of photosynthetic capacity \citep{quebbeman2016optimal}. For this dataset the correlation between \(V_{cmax25}\) and \(J_{max25}\) was $r^2=0.53$ for both cowpea and bean, which is higher than the optional \(r\) penalty (penalized when \(r^2<0.49\)). For cowpea the ratio \(J_{max25}\):\(V_{cmax25}\) was 2.02, and was 2.03 for bean. \cite{wullschleger1993biochemical} reported an average ratio of 2.05 across a set of annual species, and \cite{medlyn2002} reported an average of 1.67 across a range of plant functional types, including a ratio of 2.4 in soybean. Their previously reported ratios demonstrate that the ratio \(J_{max25}\):\(V_{cmax25}\) we extracted is reasonable.

\begin{figure}
\centering
\includegraphics[width=\textwidth]{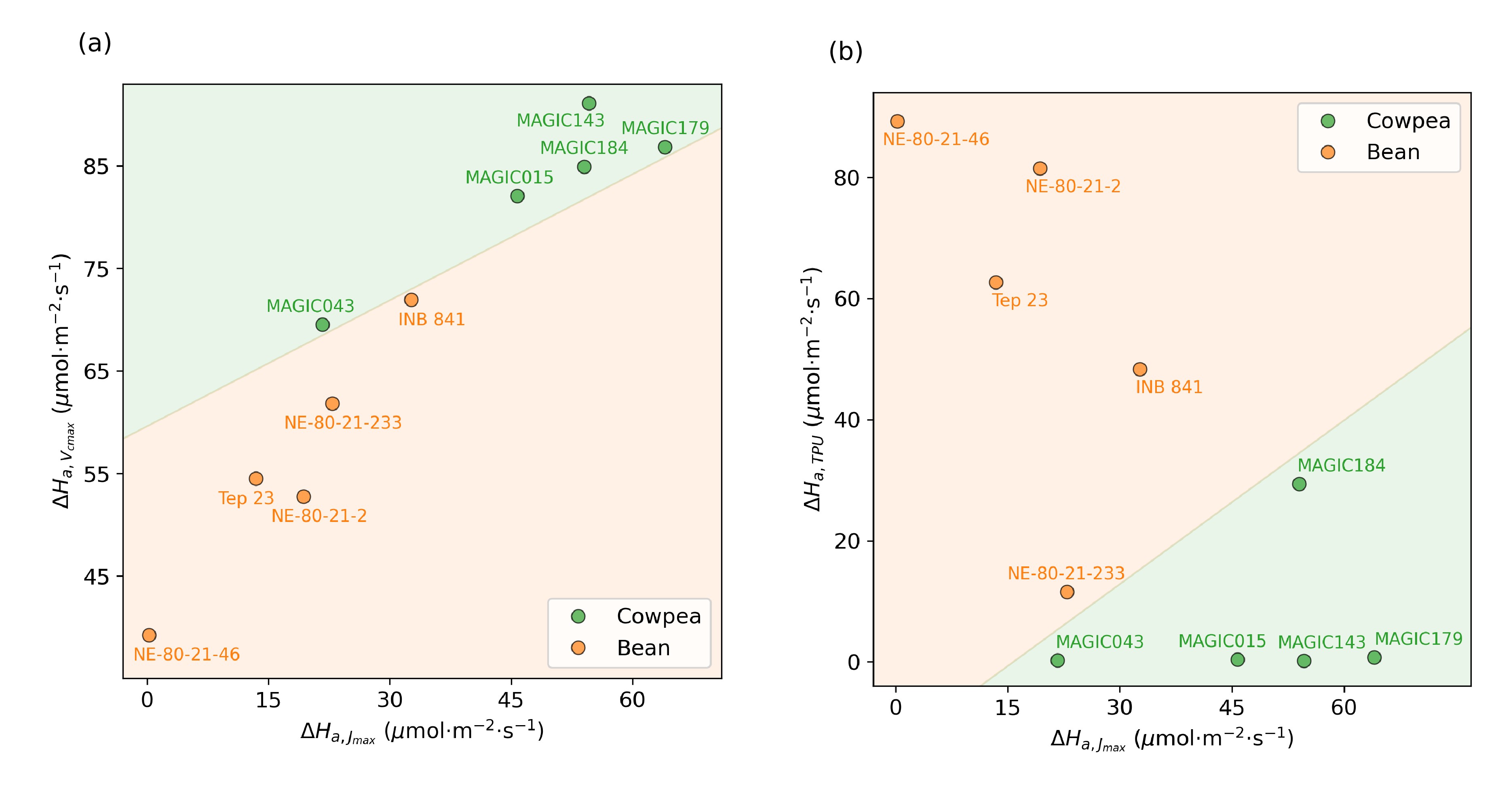}
\caption{Correlation between temperature response parameters for each genotype in the bean and cowpea datasets. (a) Fitted \(\Delta{H_{a,J_{max}}}\) versus fitted  \(\Delta{H_{a,V_{cmax}}}\). (b) Fitted \(\Delta{H_{a,J_{max}}}\) versus fitted  \(\Delta{H_{a,\mathrm{TPU}}}\). Colored zones delineate bean and cowpea genotypes.}
\label{fig:tp1dha}
\end{figure}

The fitted \(\Delta{H_a}\) parameters displayed in Fig. \ref{fig:tp1dha} are indicative of the response of net photosynthesis to temperature, and could be used as a breeding trait for selecting genotypes with high or low sensitivity to temperature. The temperature responses of Rubisco-limited and electron transport-limited carboxylation rates, as represented by \(\Delta{H_{a,V_{cmax}}}\) and  \(\Delta{H_{a,J_{max}}}\), appear positively correlated, with clear separation between cowpea and bean species (Fig. \ref{fig:tp1dha}). Bean exhibits higher photosynthetic temperature optima than cowpea, however, the mean of the maximum \(A\) of each curve of bean and cowpea are 37.41 and 45.17 \(\mu\)mol/m\(^2\)/s, respectively, suggesting a trade-off between maximum photosynthetic capacity and heat tolerance. Our fitting results suggest that the temperature response of the triose-phosphate limited pathway, \(\Delta{H_{a,\mathrm{TPU}}}\), however, is negatively correlated to that of the above two pathways, which is consistent with the findings of \cite{kumarathunge2019no}. 

\subsection{Fitting non-steady-state \(A/C_i\) curves with steady-state light response}

\begin{figure}
\centering
\includegraphics[width=10cm]{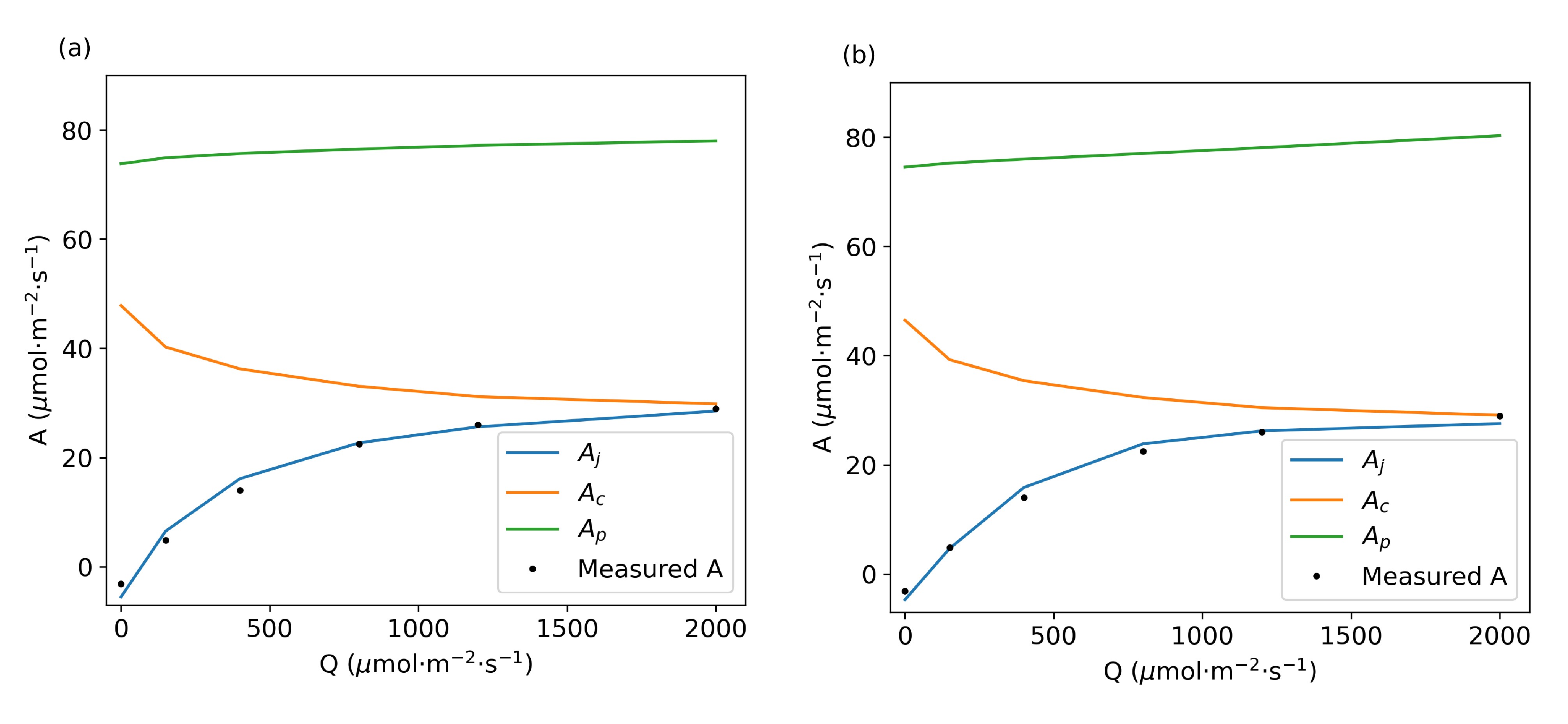}
\caption{Fitted \(A_c/Q\), \(A_j/Q\), and \(A_p/Q\) curves, and measured \(A/Q\) (black dot points) for light response type 1 (a) and 2(b). Curves in (a) and (b) are the same as curves from Fig. \ref{fig:acil1}j and \ref{fig:acil2}j, respectively.}
\label{fig:aql12}
\end{figure}

The present optimizer can also simultaneously fit \(A/C_i\) curves and light response curves (\(A/Q\)). Figures \ref{fig:acil1} and \ref{fig:acil2} show fitted \(A_c\), \(A_j\), and \(A_p\) and measured \(A\) including the light response curves (Fig. \ref{fig:acil1}j and \ref{fig:acil2}j). The \(A_p\) in both light response plot in Fig. \ref{fig:acil1} and \ref{fig:acil2} are not updated, which prevents failure of fitting. All fitting results using both light response types 1 and 2 had very high goodness of fit and appear visually reasonable. The goodness of fit for the light response curve in Fig. \ref{fig:acil2}j (non-rectangular hyperbola) is better than that of Fig. \ref{fig:acil1}j (rectangular hyperbola). This result is reasonable since the light response type 2 has more degrees of freedom due to its two parameters, $\alpha$ and $\theta$, whereas light response type 1 only has one, $\alpha$. The \(A/Q\) curves demonstrate clearer fitting results for two light response cases (Fig. \ref{fig:aql12}). These results indicate that the present optimizer can fit \(A/C_i\) curves and light response curves (\(A/Q\)) simultaneously with high accuracy.

\subsection{Comparison of non-steady-state \(A/C_i\) fitting with plantecophys}

As it is one of the most popular \(C_3\) plant \(A/C_i\) curve fitting packages, plantecophys was used as a benchmark against which PhoTorch could be compared. Table \ref{tab:compare} shows the \(\mathrm{RMSE}\) results obtained using plantecophys and PhoTorch with and without the positive \(R_{d}\) constraint. For the \(A/C_i\) curves considered in this test, the overall goodness of fit between plantecophys and PhoTorch was similar. In some cases, plantecophys was able to achieve an artificially good fit (mean \(\mathrm{RMSE}=0.4639\) \(\mu\)mol/m\(^2\)/s) by allowing \(R_{d}\) to be negative. However, if the positive constraint was removed in PhoTorch, the mean \(\mathrm{RMSE}\) (0.4556) for PhoTorch was lower than plantecophys. These tests are limited by the fact that plantecophys cannot simultaneously fit light and temperature response, so it was not possible to benchmark these more complicated fitting cases.

A 12th Gen Intel Core i9-12950HX CPU was used to run all curve fits in Table \ref{tab:compare}. Results illustrate that PhoTorch was more than four times faster than plantecophys in performing the model fitting, which could be significant if processing a large amount of unsteady \(A/C_i\) curves that contain hundreds of data points each.

\begin{table}
\centering
\caption{\(\mathrm{RMSE}\) results obtained by plantecophys and PhoTorch for fitting \(A/C_i\) curves of cowpea MAGIC043. \(\mathrm{RMSE}\) values have units of $\mu$mol/m$^2$/s.}
\begin{tabular}{>{\raggedright\arraybackslash}p{2.4cm} >{\raggedright\arraybackslash}p{2.8cm}>{\raggedright\arraybackslash}p{2.8cm}>{\raggedright\arraybackslash}p{2.8cm}}
  \hline
  $^a$Sample ID & $^b$plantecophys & $^c$PhoTorch (+$R_d$) & $^d$PhoTorch ($\pm R_d$) \\
  \hline
    5 & 0.7519 & 0.5746 & 0.4705 \\
    7 & 0.4497 & 0.5134 & 0.5443 \\
    8 & 0.6280 & 0.5234 & 0.4221 \\
    41 & 0.3274 & 0.5130 & 0.3368 \\
    43 & 0.4221 & 0.4298 & 0.4341 \\
    48 & 0.4954 & 0.5318 & 0.5436 \\
    51 & 0.3617 & 0.4371 & 0.3929 \\
    102 & 0.2737 & 0.4027 & 0.3118 \\
    105 & 0.3632 & 0.5535 & 0.4624 \\
    108 & 0.3478 & 0.4383 & 0.3761 \\
    114 & 0.6821 & 0.8783 & 0.7170 \\
  \hline
    Mean & 0.4639 & 0.5269  & 0.4556 \\
    Fitting time & 287s & 64s & 64s \\
  \hline
\end{tabular}
\parbox{\textwidth}{
\small {$^a$: Sample IDs are provided by the original dataset. \par
$^b$: The ``fitacis" funtion was used and its option ``fitTPU" was set to ``True", ``EdVC" (\(\Delta{H_{d,V_{cmax}}}\)) was set to 0,``EdVJ" (\(\Delta{H_{d,J_{max}}}\)) was set to 0,``Tcorrect" was set to ``FALSE"; other settings were set to default. It should be noted that \(R_d\) does not have positive constraint in plantecophys. \par
$^c$: The temperature and light response types were set to 0, and \(K_{c25}\), \(K_{o25}\), and \(\Gamma_{25}^*\) were not fitted. \(R_d\) has positive constraint. The learning rate and maximum iteration were set to 0.08 and 20,000, respectively. \par
$^d$:  \(R_d\) did not have positive constraint, other settings were same with $^c$.}}
\label{tab:compare}
\end{table}

\begin{figure}
\centering
\includegraphics[width=\textwidth]{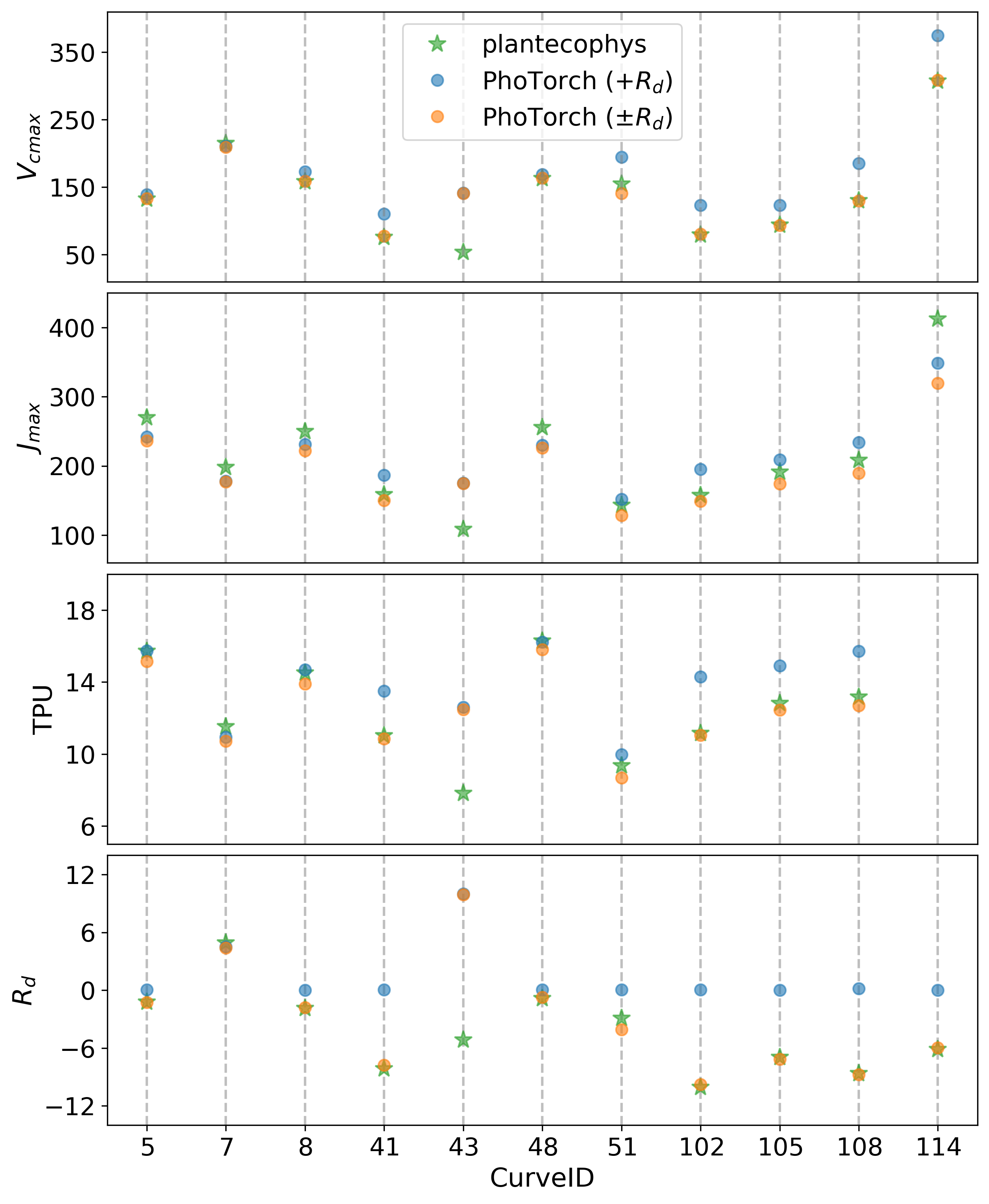}
\caption{Fitted \(V_{cmax}\), \(J_{max}\), \(TPU\), and \(R_d\) parameters for all samples of cowpea MAGIC043 using plantecophys (green star), PhoTorch with a positive \(R_d\) constraint (blue circle), and PhoTorch without an \(R_d\) constraint (orange circle).}
\label{fig:compparams}
\end{figure}

Figure \ref{fig:compparams} displays the fitted \(V_{cmax}\), \(J_{max}\), \(TPU\), and \(R_d\) parameters for all samples of cowpea MAGIC043 using plantecophys, PhoTorch with a positive \(R_d\) constraint, and PhoTorch without a \(R_d\) constraint. Generally, PhoTorch without a positive \(R_d\) constraint produced parameter estimates similar to those fitted by plantecophys. However, compared to the parameters fitted by PhoTorch with a positive \(R_d\) constraint, \(V_{cmax}\), \(J_{max}\), and \(TPU\) were underestimated. The parameters for curve 43 (Table \ref{tab:compare}) show the most significant difference between plantecophys and the two PhoTorch fittings. Although the ``true" parameters are unknown, the \(V_{cmax}\), \(J_{max}\), and \(TPU\) fitted by plantecophys are the lowest among all, which may indicate that the plantecophys fitting result is not reliable for this curve.  

\subsection{Fitting synthetic \(A/C_i\) curves}

Table \ref{tab:synresults} presents the \(\mathrm{RMSE}\) results of parameter estimation from synthetic \(A/C_i\) curves using PhoTorch. As expected, the fitted parameters are closer to the known parameters when the noise level is lower. Even at high noise levels, the fitted parameters remain reasonable, demonstrating the robustness of PhoTorch. It is worth noting that a standard deviation of 1.5 represents a significant amount of noise, with 95\% of the data points deviating between -3 and 3 $\mu$mol/m$^2$/s around the original curve. The noise especially impacts the Rubisco-limited state (\(V_{cmax}\)), as the \(A\) values in this state are much lower than than in other states, and thus the magnitude of the noise is closer to the magnitude of \(A\) itself. This explains why the fitting results for \(V_{cmax}\) have the highest \(\mathrm{RMSE}\). When using the plantecophys to fit the synthetic \(A/C_i\) curve with a 1.5 standard deviation noise, the results of  \(J_{max}\) (\(>\)40) and \(R_d\) (\(>\)1) are not satisfactory. The main reason is that the plantecophys lacks a non-negative constraint for \(R_d\).
  
\begin{table}
\centering
\caption{\(\mathrm{RMSE}\) results of fitting synthetic \(A/C_i\) curves with various standard deviations of Gaussian noise. All values have units of $\mu$mol/m$^2$/s.}
\begin{tabular}{>{\raggedright\arraybackslash}p{2.6cm} >{\raggedright\arraybackslash}p{2.2cm}>{\raggedright\arraybackslash}p{2.2cm}>{\raggedright\arraybackslash}p{2.2cm}}
  \hline
  Standard deviation of Gaussian noise &  \(V_{cmax}\) \(\mathrm{RMSE}\) (mean: 181.38 \(\mu\)mol/m\(^2\)/s) & \(J_{max}\) \(\mathrm{RMSE}\) (mean: 216.02 \(\mu\)mol/m\(^2\)/s) &  \(R_{d}\) \(\mathrm{RMSE}\) (mean: 1.34 \(\mu\)mol/m\(^2\)/s) \\
  \hline
  1.5 & 6.6208 & 6.3433 & 0.6874 \\
  1 & 3.8251 & 3.2686 & 0.6657 \\
  0.5 & 2.1196 & 1.2249 & 0.2831\\
  0.2 & 1.4774 & 1.3708 & 0.2691\\
  0 & 0.5329 & 0.3463 & 0.0717\\

  \hline
\end{tabular}
\parbox{\textwidth}{
\small {Note: Since \(\mathrm{TPU}\) limitation states did not appear in some curves, the \(\mathrm{TPU}\) fitting results were not compared. No pre-processing procedure was applied to the noisy curves.}}
\label{tab:synresults}
\end{table}

\subsection{Fitting all available parameters for both steady-state and non-steady-state \(A/C_i\) curves}

Table \ref{tab:paramall} presents fitting results for the oak tree, cowpea, and bean datasets, where all possible parameters were fitted except for \(g_m\) and light response parameters (see Table \ref{tab:paramall}). The cowpea and bean curves consist of all \(A/C_i\) curves presented in Table \ref{tab:data} except the light response curve. Each curve has its own fitted \(V_{cmax25}\), \(J_{max25}\), \(\mathrm{TPU_{25}}\), and \(R_{d25}\) value, while all other parameters listed in Table \ref{tab:paramall} are shared within the same species. Low \(\mathrm{RMSE}\) (\(<\)0.5 \(\mu\)mol/m\(^2\)/s) and high \(R^2\) (\(>\)0.98) values were obtained for both steady-state and non-steady-state \(A/C_i\) curves, demonstrating the flexibility and generality of PhoTorch when fitting a large number of parameters. While the cowpea and bean datasets contain a wide range of temperature conditions that exceed the photosynthetic temperature optimum, it should be noted that the temperature response parameters for oak may be unreliable due to the small measured temperature range in the dataset (22-28\degree{C}). 

\begin{table}
\centering
\caption{Fitted parameters for datasets from 3 species, containing both steady-state and non-steady-state \(A/C_i\) curves. All possible parameters were fitted except for the mesophyll conductance \(g_m\) and the light response parameters (the oak dataset lacked this data). The fitting produced a unique \(V_{cmax25}\), \(J_{max25}\), \(\mathrm{TPU_{25}}\), and \(R_{d25}\) for each \(A/C_i\) curve (not reported), whereas each value reported in the table is the value fitted across all curves in each species. Error metrics were calculated for the fit to each curve, and the reported value is the mean value across all curves.}
\begin{tabular}{>{\raggedright\arraybackslash}p{3.7cm} >{\raggedright\arraybackslash}p{3.2cm}>{\raggedright\arraybackslash}p{2.8cm}>{\raggedright\arraybackslash}p{2.8cm}}
  \hline
  Parameter & Oak & Cowpea & Bean \\
  \hline
  Mean \(\mathrm{RMSE}\) (\(\mu\)mol/m\(^2\)/s)  & 0.4919 & 0.4784 & 0.4649 \\
  Mean \(R^2\)  & 0.9884 & 0.9950 & 0.9873 \\
  \(K_{c25} \) & 450.7 & 322.8 & 364.9\\
  \(K_{o25} \) & 220.9 & 360.8 & 320.2\\
  \(\Gamma_{25}^* \) & 42.95 & 48.40 & 43.44\\
  \(\Delta{H_{a,V_{cmax}}}\) & 64.54 & 64.22 & 60.59\\
  \(\Delta{H_{a,J_{max}}}\) & 82.41 & 57.51 & 30.49\\
  \(\Delta{H_{a,\mathrm{TPU}}}\) & 47.48 & 15.71 & 48.50\\
  \(T_{opt,V_{cmax}}\) & 293.8 & 310.5 & 313.3\\
  \(T_{opt,J_{max}}\) & 298.5 & 311.2 & 312.6\\
  \(T_{opt,\mathrm{TPU}}\) & 301.0 & 302.2 & 313.3\\
  \(\alpha_g\) & 1.4574 & 0.1280 & 0.0546 \\
  Instrument & LI-6400 \& LI-6800 & LI-6800 & LI-6800 \\
  State of response & Steady-state & *Dynamic & Dynamic \\
  Number of curves & 41 & 55 & 60 \\
  Number of points & 566 & 9410 & 10154 \\
  \hline
\end{tabular}
\parbox{\textwidth}{
\small {*: ``Dynamic" refers to Dynamic Assimilation™ Technique feature of the LI-6800, which is a non-steady-state measurement.}}
\label{tab:paramall}
\end{table}

\section{Discussion}

Novel FvCB model fitting software was developed based on optimizers within the PyTorch deep learning framework, which was demonstrated to be able to flexibly and efficiently fit to a range of leaf-level photosynthesis CO\textsubscript{2}, light, and temperature response data without any \textit{ad hoc} parameter tuning. The resulting set of fitted parameters quantifies photosynthetic response to these environmental factors, which can benefit plant ecophysiological research. For instance, \cite{kumarathunge2019acclimation} empirically found the ratio of \(J_{max25}\) to \(V_{cmax25}\) to be negatively correlated with the plant's acclimation temperature as well as evolutionarily-adapted temperature. Theoretical arguments and empirical observations of \cite{quebbeman2016optimal} suggest that \(J_{max25}\):\(V_{cmax25}\) should increase linearly with nitrogen content, decrease exponentially with maximum daily irradiance, and be able to vary on a monthly timescale. These findings highlight the importance of robust photosynthetic parameter estimation for understanding plant responses to varying environmental conditions, which the PhoTorch software can provide. 

Leveraging of the PyTorch framework, originally developed for machine learning applications, facilitated robust parameter optimization with high computational efficiency, and the ability to enforce a number of constraints that ensure robustness in the presence of unusual or noisy gas exchange data. In addition, as the FvCB model was written entirely in PyTorch, it can be easily integrated with other AI models used in photosynthesis research. This feature is especially beneficial for expanding typical phenotyping studies focused on predicting photosynthesis traits from indirect measurements using machine learning models \citep{camino2022detecting,deng2024estimation}. These studies seek to predict FvCB parameters from sensing data such as leaf reflectance spectra or images, which relies on parameters fitted from ground truth \(A/C_i\) curves, making the entire workflow a two-stage process (first fitting \(A/C_i\) curves, then predicting the fitted parameters using sensing data). However, the present package allows for directly connecting the FvCB model with deep learning models, integrating them into a unified deep learning model. This enables the direct fitting of \(A/C_i\) curves using sensing data, which could potentially improve the reliability of phenotyping models.
 
\section*{Declaration of Competing Interest}
The authors declare that they have no known competing financial interests or personal relationships that could have appeared to influence the work reported in this paper.

\section*{Acknowledgements}
This work was supported, in whole or in part, by the Bill \& Melinda Gates Foundation INV-0028630. Under the grant conditions of the Foundation, a Creative Commons Attribution 4.0 Generic License has already been assigned to the Author Accepted Manuscript version that might arise from this submission. We acknowledge Wynn Vonnegut and Katie Risoen for their assistance in collecting gas exchange data in 2022. We thank Christine Diepenbrock, Sassoum Lo, and Jonathan Berlingeri for design and execution of the bean and cowpea field experiments from which gas exchange data was collected. We gratefully acknowledge Santos Barrera Lemus and team for their role in making the crosses for the NE lines studied herein, and Carlos Urrea and team for providing seed for these three lines and INB 841. We gratefully acknowledge Tim Porch and team for providing seed of TARS-Tep 23, and Bao Lam Huynh and Philip Roberts for providing seed of the five cowpea MAGIC lines studied herein.

\appendix
\gdef\thesection{Appendix \Alph{section}}
\setcounter{table}{0}

\section{Pedigrees of common/tepary bean used}\label{App:pedigrees}

\begin{table}[h!]
\centering
\caption{Pedigree of bean and cowpea genotypes used for field data collection.}
\begin{tabular}{>{\raggedright\arraybackslash}p{2.4cm} >{\raggedright\arraybackslash}p{8.5cm} >{\raggedright\arraybackslash}p{2.2cm}}
  \hline
  \textbf{Line ID} & \textbf{Pedigree}\\
  \hline
  NE-80-21-2 & (VAP 1xG 40019)F1 X SEF 10/-001F1-001C-02C-MC-MC-02L-MSB \\
  NE-80-21-46 & ((VAP 1xG 40019)F1 X SEF 10)F1 X SMC 214/-002F1-04C-MC-MC-01L-MSB \\
  NE-80-21-233 & (((((INB 834xG 40264)F1 X INB 834)F1 X INB 841)-003F1 X G 40036)F1 X ICTA LIGERO)F1 X SEF 10/-004F1-02C-MC-MC-01L-MSB \\
  TARS-Tep 23 & PI 502217-s x PI 440799 \\
  INB 841  & INB108 x INB605 \\
  MAGIC015  & [(CB27 x IT82E-18) x (IT89KD-288 x IT84S-2049)] x [(Suvita2 x IT00K-1263) x (IT84S-2246 x IT93K-503-1)] \\
  MAGIC043  & [(CB27 x IT82E-18) x (IT89KD-288 x IT84S-2049)] x [(Suvita2 x IT00K-1263) x (IT84S-2246 x IT93K-503-1)] \\
  MAGIC143  & [(CB27 x IT82E-18) x (IT89KD-288 x IT84S-2049)] x [(IT84S-2246 x IT93K-503-1) x (Suvita2 x IT00K-1263)] \\
  MAGIC179   & [(CB27 x IT82E-18) x (IT89KD-288 x IT84S-2049)] x [(Suvita2 x IT00K-1263) x (IT84S-2246 x IT93K-503-1)] \\
  MAGIC184  & [(CB27 x IT82E-18) x (IT89KD-288 x IT84S-2049)] x [(Suvita2 x IT00K-1263) x (IT84S-2246 x IT93K-503-1)] \\
  \hline
\end{tabular}
\label{tab:pedigrees}
\end{table}

\section{Fitting results supplementary figures}\label{App:fitting results}

\begin{figure}
\centering
\includegraphics[width=\textwidth]{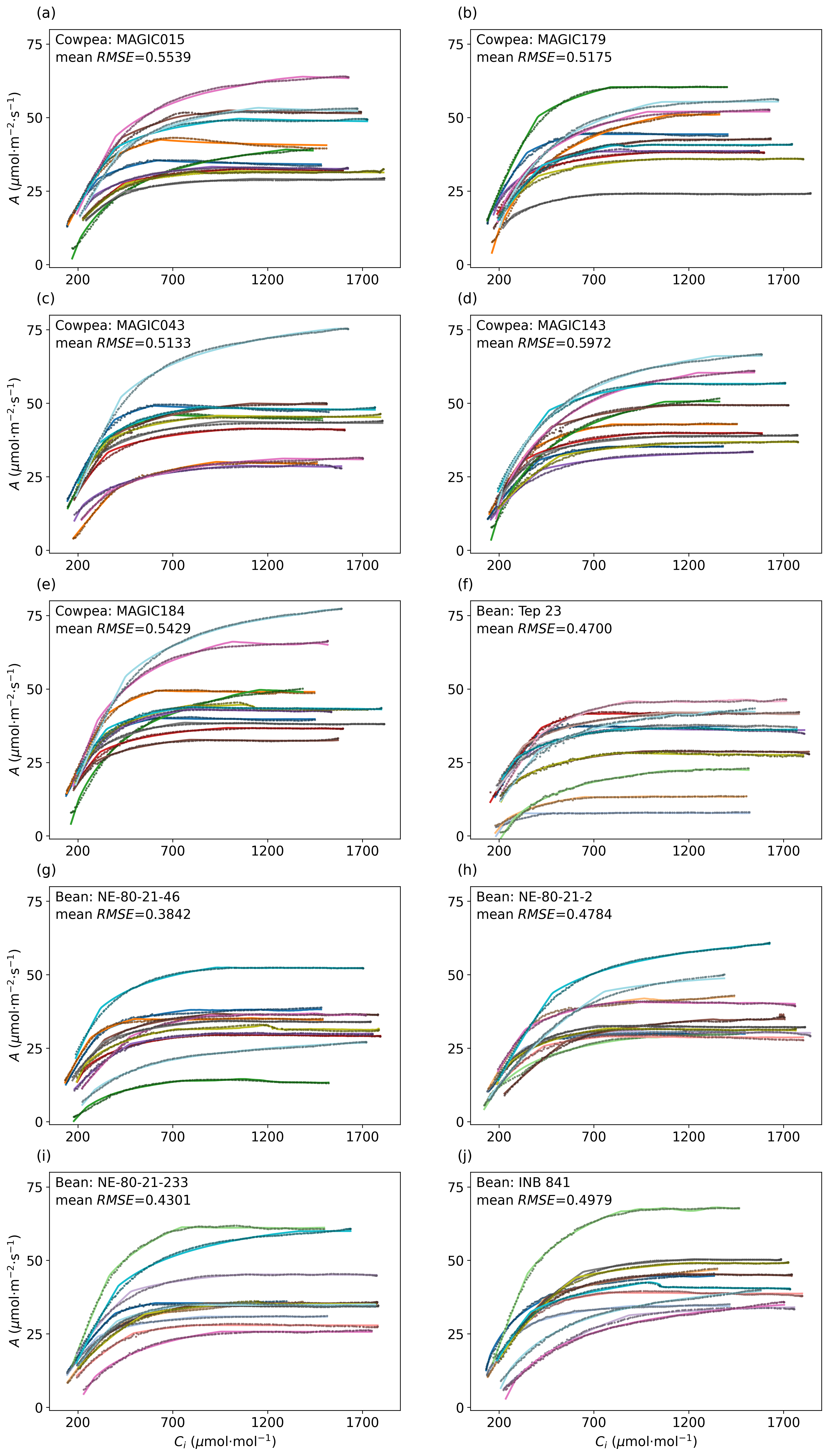}
\caption{Results of \(A/C_i\) fitting based on bean and cowpea field gas exchange data, excluding the light response curve. The dotted points are measured \(A/C_i\) values and solid curves are \(A/C_i\) curves produced by the fitted FvCB model. The learning rate and maximum iterations were set to 0.08 and 20,000 for all fittings.}
\label{fig:acifit}
\end{figure}

\begin{figure}
\centering
\includegraphics[width=15.3cm]{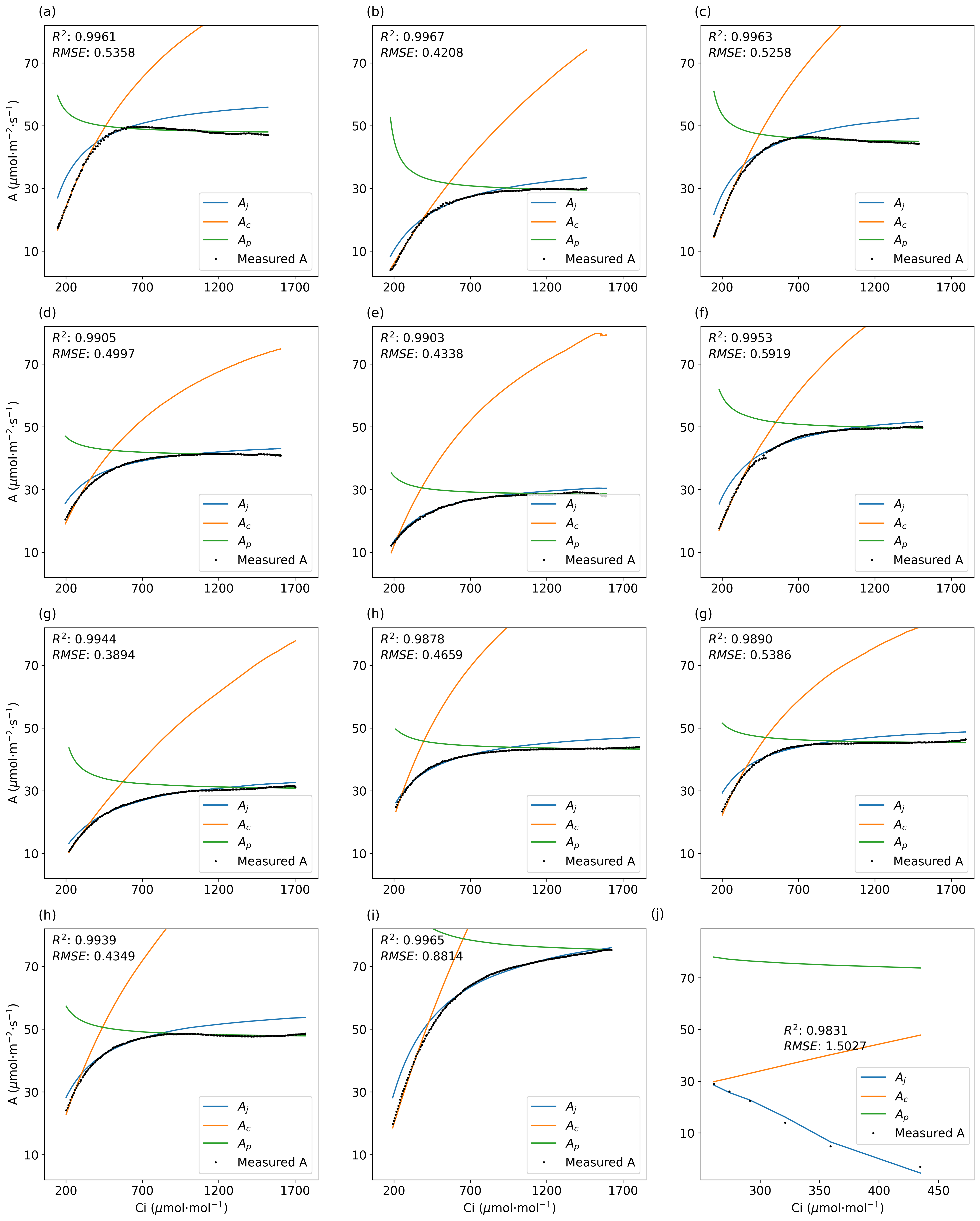}
\caption{Fitted \(A_c\), \(A_j\), and \(A_p\) curves, and measured \(A\) values (black dot points) for cowpea of genotype MAGIC043. Selected light response type is 1 (learnable parameter is $\alpha$ in Eq. \ref{eq:lr1}). The learning rate and maximum iteration were set to 0.08 and 20,000 for all fittings. Each plot has its own set of main four parameters, but all plots share the same set of temperature and light response parameters.}
\label{fig:acil1}
\end{figure}

\begin{figure}
\centering
\includegraphics[width=15.3cm]{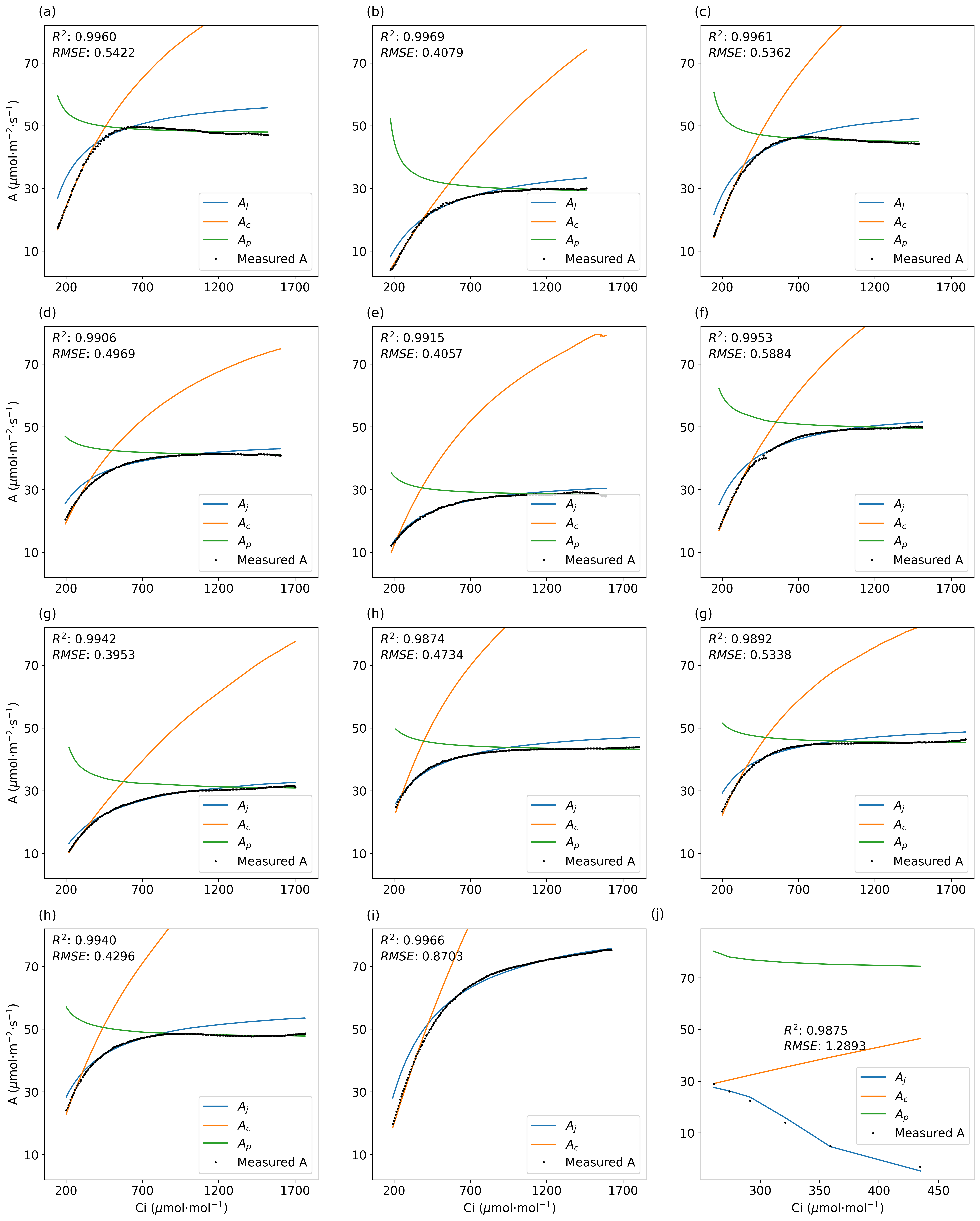}
\caption{Fitted \(A_c/C_i\), \(A_j/C_i\), and \(A_p/C_i\) curves, and measured \(A/C_i\) values (black dot points) for cowpea of genotype MAGIC043. Selected light response type is 2 (learnable parameters are $\alpha$ and $\theta$ in Eq. \ref{eq:lr1}). The learning rate and maximum iteration were set to 0.08 and 20,000 for all fittings.}
\label{fig:acil2}
\end{figure}

\bibliographystyle{elsarticle-harv}

\bibliography{Ref}
\end{document}